%% file: main-sbrc26.tex
\documentclass[12pt]{article}

\input{pacotes}
\input{preambulo}

\newcommand{\alb}[1] {\textcolor{red}{ALB: #1}}
\newcommand{\msn}[1] {\textcolor{blue}{MSN: #1}}

\input{titulo}
\begin{document}
	
\maketitle
\input{abstract}
\input{resumo}
\input{introducao}
\input{fundamentacao}
\input{recall-rcl}

\input{rcl-solidity}

\input{conclusao}

\bibliographystyle{sbc}
\bibliography{refs}
\end{document}

%% file: pacotes.tex
\usepackage{sbc-template}
\usepackage[utf8]{inputenc}
\usepackage[T1]{fontenc}
\usepackage[brazil]{babel}

\usepackage{graphicx}
\usepackage{url}

\usepackage{amssymb}
\usepackage{amsmath}

\usepackage{comment}
\usepackage{lipsum}
\usepackage{framed}

\usepackage[table,xcdraw]{xcolor}

\usepackage{tikz}
\usetikzlibrary{petri, positioning}

\usepackage{algorithm}
\usepackage{algpseudocode}
\usepackage{listings}



%% file: preambulo.tex
\definecolor{verylightgray}{rgb}{.97,.97,.97}

\lstdefinelanguage{Solidity}{
    keywords=[1]{anonymous, assembly, assert, balance, break, call, callcode, case, catch, class, constant, continue, constructor, contract, debugger, default, delegatecall, delete, do, else, emit, event, experimental, export, external, false, finally, for, function, gas, if, implements, import, in, indexed, instanceof, interface, internal, is, length, library, log0, log1, log2, log3, log4, memory, modifier, new, payable, pragma, private, protected, public, pure, push, require, return, returns, revert, selfdestruct, send, solidity, storage, struct, suicide, super, switch, then, this, throw, transfer, true, try, typeof, using, value, view, while, with, addmod, ecrecover, keccak256, mulmod, ripemd160, sha256, sha3}, 
    keywordstyle=[1]\color{blue}\bfseries,
    keywords=[2]{address, bool, byte, bytes, bytes1, bytes2, bytes3, bytes4, bytes5, bytes6, bytes7, bytes8, bytes9, bytes10, bytes11, bytes12, bytes13, bytes14, bytes15, bytes16, bytes17, bytes18, bytes19, bytes20, bytes21, bytes22, bytes23, bytes24, bytes25, bytes26, bytes27, bytes28, bytes29, bytes30, bytes31, bytes32, enum, int, int8, int16, int24, int32, int40, int48, int56, int64, int72, int80, int88, int96, int104, int112, int120, int128, int136, int144, int152, int160, int168, int176, int184, int192, int200, int208, int216, int224, int232, int240, int248, int256, mapping, string, uint, uint8, uint16, uint24, uint32, uint40, uint48, uint56, uint64, uint72, uint80, uint88, uint96, uint104, uint112, uint120, uint128, uint136, uint144, uint152, uint160, uint168, uint176, uint184, uint192, uint200, uint208, uint216, uint224, uint232, uint240, uint248, uint256, var, void, ether, finney, szabo, wei, days, hours, minutes, seconds, weeks, years},	
    keywordstyle=[2]\color{teal}\bfseries,
    keywords=[3]{block, blockhash, coinbase, difficulty, gaslimit, number, timestamp, msg, data, gas, sender, sig, value, now, tx, gasprice, origin},	
    keywordstyle=[3]\color{violet}\bfseries,
    identifierstyle=\color{black},
    sensitive=true,
    comment=[l]{//},
    morecomment=[s]{/*}{*/},
    commentstyle=\color{gray}\ttfamily,
    stringstyle=\color{red}\ttfamily,
    morestring=[b]',
    morestring=[b]"
}

\tikzset{
	place/.style={
		circle, draw,
		align=center,
		minimum size=2mm,
		text width=1.7cm, 
		align=center,
		inner sep=1pt,
		font=\scriptsize
	},
	place_maquina_lavar/.style={ 
		circle, draw,
		align=center,
		minimum size=8mm,
		text width=1.7cm, 
		align=center,
		inner sep=1pt,
		font=\scriptsize
	},
	transition/.style={
		rectangle, draw,
		minimum height=3mm,
		minimum width=5mm,
		align=center,
		font=\scriptsize
	},
	place_compacto/.style={
		circle, draw, align=center, font=\scriptsize,
		minimum size=1.7cm, inner sep=1pt 
	},
	cliente_compacto/.style={
		place_compacto, draw=blue, text=blue
	},
	comanda_compacto/.style={
		place_compacto, draw=red, text=red
	},
	prato_compacto/.style={
		place_compacto, draw=olive, text=olive
	},
	trans_compacto/.style={
		transition, align=center
	},
	token/.style={
		circle, draw,
		align=center,
		minimum size=2mm,
		text width=0.6cm, 
		align=center,
		inner sep=1pt,
		fill=green,
		font=\scriptsize
	},
	cliente/.style={place, draw=blue, text=blue,minimum size=4mm},
	comanda/.style={place, draw=red, text=red, minimum size=6mm},
	prato/.style={place, draw=olive, text=olive, minimum size=4mm}
}


%% file: titulo.tex
\title{Uma Prova de Conceito para a Verificação Formal de Contratos Inteligentes}

\author{Murilo de Souza Neves\inst{1}, Adilson Luiz Bonifacio\inst{1} }

\address{Departamento de Computação -- Universidade Estadual de Londrina
	(UEL)\\
	Caixa Postal 10.011 -- CEP 86.057-970 -- Londrina -- PR -- Brazil
	\email{\{murilo.souza.neves,bonifacio\}@uel.br}
}

%% file: abstract.tex
\begin{abstract}
Smart contracts are tools with self-execution capabilities that provide enhanced security compared to traditional contracts; however, their immutability makes post-deployment fault correction extremely complex, highlighting the need for a verification layer prior to this stage. Although formalisms such as Contract Language (CL) enable logical analyses, they prove limited in attributing responsibilities within complex multilateral scenarios. This work presents a proof of concept using the Relativized Contract Language (RCL) and the RECALL tool for the specification and verification of a purchase and sale contract involving multiple agents. The study demonstrates the tool's capability to detect normative conflicts during the modeling phase. After correcting logical inconsistencies, the contract was translated into Solidity and functionally validated within the Remix IDE environment, confirming that prior formal verification is fundamental to ensuring the reliability and security of the final code.
\end{abstract}

%% file: resumo.tex
\begin{resumo} 
Contratos Inteligentes são ferramentas com capacidade de autoexecução que fornecem uma maior segurança se comparados aos contratos comuns, entretanto a sua imutabilidade torna a correção de falhas após a implantação extremamente complicada, evidenciando a necessidade de uma camada de verificação anterior a essa etapa. Embora formalismos como a Linguagem de Contratos (CL) permitam análises lógicas, estes mostram-se limitados na atribuição de responsabilidades em cenários multilaterais complexos. Este trabalho apresenta uma prova de conceito utilizando a Lógica de Contratos Relativizada (RCL) e a ferramenta RECALL para a especificação e verificação de um contrato de compra e venda envolvendo múltiplos agentes. O estudo demonstra a capacidade da ferramenta em detectar conflitos normativos ainda na fase de modelagem. Após a correção das inconsistências lógicas, o contrato foi traduzido para a linguagem Solidity e validado funcionalmente no ambiente Remix IDE, confirmando que a verificação formal prévia é fundamental para garantir a confiabilidade e a segurança do código final.
Tópicos
\end{resumo}

%% file: introducao.tex
\section{Introdução}

A popularização da tecnologia \emph{blockchain} e o surgimento do \emph{Ethereum} possibilitaram 
contratos autoexecutáveis, conhecidos como Contratos Inteligentes \cite{kolvart-smart-contracts}.
Estes contratos prometem eliminar intermediários e garantir a execução fiel de termos acordados, evitando possíveis ambiguidades relacionadas à linguagem natural. 
No entanto, a imutabilidade inerente às \emph{blockchains} torna qualquer erro no código crítico, uma vez que correções após a implantação são complexas e, muitas vezes, inviáveis. 
Diante desse cenário, a verificação formal de contratos torna-se um requisito fundamental para garantir a segurança e confiabilidade das transações digitais \cite{wellington-conflitos}

As técnicas de verificação têm sido aplicadas em diversas áreas, inclusive em contratos eletrônicos \cite{agarwal-contratos-eletronicos} com o objetivo de 
contornar problemas tais como ambiguidades e inconsistências. 
Formalismos como a Linguagem de Contratos (CL) \cite{fenech-CL}, baseada na lógica deôntica padrão,  permitem uma análise sistemática. 
No entanto, ainda são insuficientes em cenários de contratos multilaterais complexos, onde é necessário se determinar explicitamente os responsáveis pela execução ou violação de uma ação. 

Para solucionar essa limitação, \cite{wellington-conflitos} propuseram a Linguagem de Contrato Relativizada (RCL), uma extensão da CL que incorpora os conceitos da Lógica Deôntica Relativizada. 
A principal inovação da RCL é a capacidade de associar indivíduos a cada cláusula, especificando o responsável por uma obrigação e a contraparte que recebe essa ação. 
A RCL então permite que todas as partes de um contrato multilateral sejam identificadas corretamente, permitindo uma verificação a partir das especificidades da linguagem RCL. 
A verificação de contratos descritos em RCL pode ser realizada com a ferramenta desenvolvida, a RECALL \cite{wellington-conflitos}.

Com o objetivo de verificar a corretude de contratos inteligentes, uma abordagem consiste em descrever contratos inteligentes em RCL. 
Dessa forma, este trabalho apresenta um estudo de caso focado na aplicação da RCL como etapa de verificação prévia para o desenvolvimento de contratos inteligentes correspondentes. 
A proposta consiste em utilizar a RECALL para validar a especificação lógica de um contrato, garantindo um contrato 
livre de conflitos normativos. 
Como prova de conceito, um contrato é especificado em RCL, verificado e, posteriormente, traduzido de forma \emph{ad-hoc} para a linguagem Solidity \cite{solidity-docs}.
Os testes de comportamento são realizados no ambiente Remix IDE, demonstrando como a verificação formal prévia pode guiar a implementação correta de contratos inteligentes na plataforma Ethereum \cite{Buterin2013}.


%% file: fundamentacao.tex
\section{Verificação, Contratos e \emph{Blockchains}}\label{cap_exemplos}

Os conceitos fundamentais que baseiam o estudo desenvolvido neste trabalho estão relacionados 
a verificação de contratos eletrônicos usando a RCL com suporte da ferramenta RECALL,  
ao conceito de contratos inteligentes e \emph{blockchain}, bem como a plataforma de propósito geral \emph{Ethereum} e a linguagem de implementação Solidity. 

\input{ver-contratos}
\input{smart-contracts}

\input{plataforma}

%% file: ver-contratos.tex
\subsection{Verificação de Contratos}

Um negócio jurídico bilateral ou multilateral em que as partes envolvidas constituem, modificam ou extinguem posições jurídicas de essência ou expressão patrimonial define a noção de contrato. 
Assim, um contrato  envolve uma declaração de vontade 
entre as partes negociantes e atribuídos pelo ordenamento jurídico,  respeitados os elementos de existência, os requisitos de validade e os fatores de eficácia contidos na norma~\cite{felipe-contratos}.   

Um contrato gera obrigações para ambas as partes envolvidas, que convencionam, por consentimento recíproco,
a dar, fazer ou não fazer alguma coisa, verificando, assim, a constituição, modificação ou extinção do vínculo patrimonial~\cite{miranda-contratos}.

Com a evolução da tecnologia surgiu uma forma eletrônica de lidar com contratos legais, os chamados contratos eletrônicos ou \emph{e-contracts}. 
Ao contrário dos contratos tradicionais, concebidos de forma física, os contratos eletrônicos se utilizam das tecnologias computacionais para criar, 
negociar, armazenar e executar contratos num ambiente digital~\cite{sushma-e-contract}.

Os e-contracts são uma parte da transição de contratos tradicionais para um ambiente virtual, amplamente utilizados no cotidiano, tais como nos termos de aplicativos, cookies de websites, entre outros. 
Estes contratos possuem diversas vantagens se comparados com contratos tradicionais. 
Porém, por estarem em uma plataforma digital, devem seguir normas e regras que adaptem esses novos contratos a leis semelhantes aos contratos tradicionais.


Erros e inconsistências de um contrato mal-formado podem trazer prejuízos a uma das partes envolvidas. 
Uma forma de garantir a corretude de um contrato é através de algum processo de verificação sobre as cláusulas que representam as regras estabelecidas pelo contrato. 
Embora os conceitos e aplicações de contratos convencionais
e e-contracts apresentem diferenças, a verificação de ambos é fundamental para assegurar que nenhuma das partes envolvidas seja lesada, prevenindo possíveis conflitos e inconsistências~\cite{wellington-conflitos-outro}.

Existem várias técnicas de análise e verificação de contratos que viabilizam um processo mais rigoroso para garantir as propriedades desejadas de um contrato. 
Entre essas técnicas, destaca-se o processo de negociação para contratos legais, 
que envolve a busca por um acordo entre as partes e que seja mutuamente aceitável. 
Em geral, cada parte inicia o processo propondo uma solução que atenda aos seus interesses; 
caso uma das partes não aceite, ocorrem contrapropostas até que ambas cheguem a um consenso.
Esse processo de negociação pode se dar em contratos bilaterais ou multilaterais, podendo ou não envolver mediadores.

Com o contrato já negociado, uma outra etapa consiste na 
detecção de conflitos, uma técnica voltada para identificar e eliminar conflitos normativos em contratos. 
Um conflito normativo pode invalidar o contrato e resultar em violações. 
Por isso, a importância de uma verificação mais precisa 
para se encontrar inconsistências antes que o contrato seja executado, após a fase de negociação.

Ambos os métodos podem ser eficientes na busca por conflitos, mas ainda dependem da linguagem natural na qual os contratos são escritos. 
Um contrato em linguagem natural pode trazer problemas de ambiguidade em suas cláusulas. 
Uma forma de contornar tais problemas é descrever os contratos por meio de linguagens e especificações formais, possibilitando uma verificação sistemática ao mesmo tempo que evita inconsistências.

Com o objetivo de buscar conflitos em contratos multilaterais e 
se beneficiar dos operadores deônticos relativizados, 
Bonifacio~e~Mura~\cite{wellington-conflitos} propuseram uma extensão para a CL\footnote{do inglês, \emph{Contract Language}}, uma linguagem para contratos desenvolvida para representar formalmente contratos legais, serviços web, interfaces e protocolos de comunicação~\cite{schneider-prisacariu-CL}. 
Essa extensão, denominada RCL\footnote{do inglês, \emph{Relativized Contract Language}}, adiciona lógica deôntica relativizada à linguagem de contratos, onde os participantes são associados 
a cada ação do contrato, permitindo que as responsabilidades sejam atribuídas apenas sobre os participantes associados à ação. 
As possíveis relativizações são: 
(i) todos os participantes do contrato realizam a ação para um outro; 
(ii) um participante realiza uma ação para todos os outros; 
(iii) um participante realiza uma ação apenas a um outro~\cite{adilson-checking-conflicts}.

Com essa linguagem, contratos mais complexos podem ser descritos de forma objetiva, evitando ambiguidades e permitindo a automatização do processo de verificação formal de tais contratos.
Neste sentido, \cite{wellington-conflitos} também desenvolveram uma ferramenta para verificar contratos escritos em RCL, chamada RECALL. 
A ferramenta possibilita a detecção de conflitos em contratos multilaterais escritos em RCL. 

%% file: smart-contracts.tex
\subsection{Contratos Inteligentes}\label{smart-contracts}

Os contratos eletrônicos dependem de intermediários para sua validação e execução~\cite{szabo-smart-contracts}. 
Já os contratos inteligentes são um conjunto de promessas, especificadas em formato digital, incluindo protocolos pelos quais as partes envolvidas cumprem essas promessas. 
Essa abordagem elimina a necessidade de intermediários e garante a execução automática e segura dos termos contratuais.
Assim, os contratos são descritos por código 
mantendo a ideia de contratos em ambiente virtual, 
eliminando a necessidade de intermediários para garantir a execução dos termos acordados, reduzindo custos e agilizando os processos, além de aumentar a eficiência e transparência sobre as transações realizadas~\cite{filippi-smart-contracts}.

Um contrato inteligente é então um programa de computador capaz de tomar decisões quando determinadas precondições são atendidas~\cite{kolvart-smart-contracts}. 
A inteligência de um contrato depende da complexidade da transação programada a ser realizada, 
em alguns casos 
muito simples e executadas em segundos ou minutos, 
em outros casos 
complexas e demoradas, 
envolvendo negociações e dezenas de páginas de texto escrito com direitos e obrigações específicas, podendo levar horas ou meses para serem concluídas.

Embora os contratos inteligentes ofereçam a promessa de contratos autoexecutáveis, vários desafios legais significativos ainda persistem. 
A imutabilidade, por exemplo, é um problema recorrente, já que qualquer alteração nas cláusulas do contrato exige que um novo contrato seja criado e verificado, gerando custos adicionais~\cite{kolvart-smart-contracts}. 

A execução de um contrato inteligente ocorre de maneira automática, conforme as condições predefinidas. 
Como o contrato é codificado, essas condições devem ser estabelecidas, bem como possíveis restrições. 
Após a criação e codificação, o contrato deve ser então implantado na \emph{blockchain}, para que as operações implementadas sejam executadas de acordo com as cláusulas definidas.

Uma \emph{blockchain} pode ser definida como um banco de dados transacional globalmente distribuído~\cite{solidity-docs}, onde um usuário pode realizar a leitura de dados tais como, documentos e dados de transações.
Porém, existem transações em que todas as partes envolvidas devem estar de acordo para serem realizadas. 
Além disso, essas transações são sempre criptograficamente assinadas pelo 
criador da transação~\cite{solidity-docs}.

Uma vez que o contrato se encontra disponível, qualquer parte envolvida pode utilizar as funções do contrato, e executar uma transação na \emph{blockchain}. 
Quando as condições codificadas no contrato são atendidas, automaticamente as ações previstas devem ser executadas. 
Por exemplo, a execução de uma ação pode transferir ativos, liberar informações ou registrar dados. 

No entanto, a execução automatizada de contratos inteligentes também apresenta desafios, especialmente quando ocorrem erros no código ou quando surgem situações não previstas durante a fase de desenvolvimento. 
Vale lembrar que uma característica importante desses contratos é que, uma vez implantados na \emph{blockchain}, alterações ou correções não podem ser facilmente realizadas. 
Essa imutabilidade garante a confiança na execução do contrato,
mas também implica em riscos. 
Assim, a imutabilidade dos contratos inteligentes tem suas vantagens, como por exemplo assegurar a execução correta dos termos, mas também suas desvantagens, como por exemplo, perpetuar falhas não detectadas antes de sua implantação. 

Para evitar que a imutabilidade dos contratos não seja um problema, é necessário que exista um mecanismo confiável para se obter uma interpretação precisa dos requisitos de um contrato para sua implementação. 
Um dos mecanismos possíveis é a detecção de conflitos em contratos, usando linguagens formais e lógicas matemáticas para representar um contrato de forma precisa, evitando possíveis conflitos e inconsistências no contrato~\cite{wellington-conflitos}.

%% file: plataforma.tex
\subsection{Uma \emph{Blockchain} de Propósito Geral}

O Ethereum é uma plataforma universal de aplicações baseadas em \emph{blockchain} proposta para praticamente todo tipo de computação através de contratos inteligentes~\cite{sergei-ethereum}.
Uma das possíveis linguagens de programação para escrever contratos inteligentes para a plataforma Ethereum é a Solidity~\cite{solidity-gavin-wood}.

A Solidity é baseada no paradigma orientado a objetos, onde a representação de um contrato se assemelha a declaração de uma classe~\cite{solidity-gavin-wood}.
Além disso, a linguagem permite princípios como o de herança e polimorfismo, além dos modificadores que permitem alterar o comportamento de funções de forma declarativa~\cite{solidity-docs}. 
Esses modificadores são comumente utilizados para verificar automaticamente condições específicas antes da execução de uma função. 
O modificador não apenas reforça a segurança do contrato, mas também elimina duplicação de código.
Além dessa aplicação clássica, os modificadores podem ser usados para executar trechos de código antes ou depois da função principal ou 
pode chamar outras funções internas do contrato.

Como um contrato inteligente é baseado na relação entre duas ou mais partes, é importante entender como o código em Solidity se relaciona com os acordos entre essas partes. 
Em geral, um contrato gera obrigações entre as partes envolvidas, que convencionam, por consentimento recíproco, a dar, fazer ou não fazer alguma coisa. 
Os participantes de um contrato, por exemplo, são variáveis do tipo \emph{address} declaradas, que representam endereços no Ethereum. 
Já as obrigações são implementadas como funções que exigem ações específicas, tal como realizar um pagamento, que pode ser descrito por um modificador ou outro método. 

Cláusulas de um contrato que possuem prazos, termos de pagamento e responsabilidades podem ser expressas por meio de combinações de recursos da linguagem, como funções com modificadores, funções que mudam o valor de variáveis, entre outras. 
As punições por descumprimento de alguma cláusula podem ser implementadas por meio de restrições, baseadas no endereço da conta do punido, ou reverter funções que detectaram um descumprimento ou até mesmo banimento do usuário punido daquele contrato. 
Através desses recursos é possível então descrever praticamente qualquer acordo entre duas ou mais partes de um contrato em Solidity.

Contratos escritos em Solidity podem ser implantados e executados na ferramenta Remix IDE, um simulador para a plataforma Ethereum que fornece um ambiente virtual para desenvolvedores escreverem, depurarem e testarem seus contratos~\cite{remix-docs}. 

%% file: recall-rcl.tex
\section{Verificação Formal de Contratos Eletrônicos}\label{aplicacao-rcl}

A verificação rigorosa de contratos eletrônicos se utiliza de linguagens formais de especificação e algoritmos precisos. 
Uma possível abordagem de verificação rigorosa pode então se valer da RCL que permite uma análise precisa das cláusulas que descrevem um contrato e também a detecção de conflitos normativos~\cite{wellington-conflitos}. 
Uma ferramenta de suporte para verificação de tais contratos é a~\emph{https://recall-site.github.io}{RECALL}. 

A Subseção~\ref{sec:contratorcl-conflito} apresenta um contrato de compra e venda de produtos, bem como suas regras expressas em RCL, inicialmente, com um conflito. 
O contrato é analisado e verificado pela ferramenta RECALL para apontar o conflito existente. 
Na sequência, a Subseção~\ref{sec:contratorcl-corrigido}  apresenta o contrato corrigido e novamente verificado para garantir um contrato livre de conflitos. 

\subsection{Um Contrato em RCL}\label{sec:contratorcl-conflito}

O contrato de compra e venda abordado em~\cite{wellington-conflitos} descreve um cenário de compra digital composta por diferentes participantes: 
um comprador, que realiza a aquisição de produtos; 
um vendedor que oferece tais produtos; 
uma transportadora também faz parte do processo, com a tarefa de entregar o produto ao comprador; e 
um banco intermedeia as transações financeiras. 
Além do acordo entre as partes, também são consideradas 
regras internas dos envolvidos.
Neste cenário, a transportadora define uma regra interna segundo a qual um produto só deve ser entregue se o frete já tiver sido pago pelo vendedor. Outra regra interna é estabelecida pelo banco, que proíbe a realização de pagamentos quando da ausência de notificações requeridas para evitar fraudes.
A descrição do contrato em linguagem natural é dada como segue:
{\fontsize{10}{10}
	\begin{enumerate}
		\item O \textbf{Comprador} realiza a compra de um produto do \textbf{Vendedor}.
		\item O \textbf{Comprador} é obrigado a pagar o produto ao \textbf{Banco}.
		\item O \textbf{Banco} deve enviar a notificação sobre o pagamento do produto ao \textbf{Vendedor}.
		\item Após o \textbf{Banco} notificar o \textbf{Vendedor} sobre o pagamento, o \textbf{Vendedor} é obrigado a enviar o produto por meio de um \textbf{Transportador} e pagar os custos de envio do produto ao \textbf{Banco}.
		\item O \textbf{Transportador} deve entregar o produto ao \textbf{Comprador}.
		\item Após a entrega do produto, o \textbf{Comprador} é obrigado a informar o \textbf{Banco} sobre a entrega do produto, enquanto o \textbf{Transportador} deve notificar o \textbf{Vendedor} que o produto foi entregue ao \textbf{Comprador}.
		\item Quando o \textbf{Vendedor} é notificado sobre a entrega do produto, o \textbf{Vendedor} é obrigado a notificar o \textbf{Banco} para liberar o pagamento dos custos de envio ao \textbf{Transportador}.
		\item Quando o \textbf{Comprador} notifica o \textbf{Banco} sobre a entrega do produto, o \textbf{Banco} libera o pagamento correspondente ao \textbf{Vendedor}.
		\item O \textbf{Banco} deve pagar os custos de envio ao \textbf{Transportador} após o \textbf{Vendedor} efetuar o pagamento do valor referido.
		
		\addvspace{\medskipamount}
		\item[] \textbf{Regras Internas do Banco} 
		
		\item O \textbf{Banco} está proibido de pagar o \textbf{Vendedor} até que uma notificação adequada seja recebida do \textbf{Comprador} confirmando que o produto foi entregue.
		\item O \textbf{Banco} está proibido de liberar o pagamento dos custos de envio para o \textbf{Transportador} até que o \textbf{Vendedor} notifique o banco.
		
		\addvspace{\medskipamount}
		\item[] \textbf{Regras Internas do transportador} 
		
		\item O \textbf{Transportador} está proibido de entregar o produto até que o \textbf{Vendedor} tenha pago os custos de envio.
	\end{enumerate}
}

Com o intuito de facilitar a compreensão do contrato abordado, 
a Tabela~\ref{tab:acoes} apresenta as ações e suas respectivas descrições. 
\begin{table}[hbt]
\centering
\begin{tabular}{|l|l|}
\hline
\textbf{ação} & \textbf{descrição} \\ \hline
buyProduct & Comprar o produto \\ \hline
payProduct & Pagar o valor do produto \\ \hline
notifyProductPayment & Notificar o pagamento do produto \\ \hline
sendProduct & Enviar o produto \\ \hline
deliverProduct & Entregar o produto \\ \hline
notifyProductReceipt & Notificar o recebimento do produto \\ \hline
notifyProductDelivery & Notificar a entrega do produto \\ \hline
payShippingCosts & Pagar o frete \\ \hline
releaseShippingCosts & Liberar o pagamento do frete \\ \hline
\end{tabular}
\caption{Ações do contrato}
\label{tab:acoes}
\end{table}
O contrato de compra e venda descrito em RCL é apresentado na Figura~\ref{fig:rclerradoteste}. 
\begin{figure}[hbt]
\centering
\input{figuras/contratoRCLErrado} 
\end{figure}
O contrato é composto por quatro participantes: 
\emph{buyer} (b), o comprador; \emph{seller} (s), o vendedor; \emph{bank} (k), o intermediário para transações financeiras; e o \emph{carrier} (c), responsável pelo transporte do produto. 
Cada ação do contrato é caracterizada pela relativização do tipo um para um, onde em cada ação as partes são especificadas. 
Por exemplo, a ação \textit{buyProduct} só pode ser 
realizada do \emph{buyer} para o \emph{seller}, nessa ordem. 
Outra característica do contrato é a presença do operador deôntico de obrigação e a penalidade associada, caso uma violação ocorra. 
Por fim, as regras internas do banco e transportadora também são especificadas. 

Essa versão original do contrato foi submetida à ferramenta RECALL que, por sua vez, retornou a presença de um conflito entre as cláusulas \emph{$\{c, b\}$(deliverProduct)} e \emph{$\{c, b\}$O(deliverProduct)}. 
A análise realizada representa um conflito entre a regra interna da transportadora e o restante do contrato. 
A regra interna estabelecia que a transportadora 
poderia enviar o produto somente após o pagamento do frete. 
Por outro lado, espera-se que o produto seja entregue pela transportadora ao comprador antes que o pagamento do frete seja liberado pelo banco. 
Portanto, existe uma incompatibilidade na regra interna da transportadora, pois o pagamento do frete, realizado pelo vendedor, era esperado antes do envio do produto ao comprador. 
Além disso, a transportadora também considerava, em sua regra interna, que o pagamento do frete deveria ser realizado pelo vendedor, e não pelo banco, para então proceder à entrega do produto.

\subsection{Contrato RCL Corrigido}\label{sec:contratorcl-corrigido}

A correção do conflito apresentado na Subseção~\ref{sec:contratorcl-conflito} consiste na inclusão de uma nova cláusula indicando que o banco deveria notificar a transportadora sobre o pagamento do frete pelo vendedor. 
Outra alteração necessária ocorre na regra interna da transportadora, que agora leva em consideração a notificação do banco como garantia de pagamento do frete.

O contrato revisado, com a modificação da quinta cláusula e das regras internas da transportadora, corrige o conflito existente. 
O trecho do contrato com as cláusulas modificadas é apresentado a seguir: 
\\\textbf{[Regra do Contrato]}
\begin{enumerate}
\item[5.] O \textbf{Banco} deve notificar a \textbf{Transportadora} sobre o pagamento do frete e após o \textbf{Banco} atestar o pagamento, a \textbf{Transportadora} é obrigada a entregar o produto para o \textbf{Comprador}.
\end{enumerate}

\textbf{[Regra interna da Transportadora]}
\begin{enumerate}
\item[12.] A \textbf{Transportadora} é proibida de entregar o produto até que o \textbf{Banco} notifique-a
de que o \textbf{Vendedor} pagou o valor do frete.
\end{enumerate}

O trecho do contrato em RCL modificado é descrito 
na Figura~\ref{fig:rclcorretoteste}. 

\begin{figure}[hbt]
\centering
\input{figuras/contratoRCLCorreto} 
\end{figure}
Após nova submissão do 
contrato corrigido à ferramenta RECALL, 
a análise obtida retornou um contrato livre de conflitos. 

Observa-se que ferramentas como a RECALL são de suma importância para assegurar contratos livres de conflitos e garantir sua corretude. 
Um caminho para a verificação de contratos inteligentes pode ser através do uso de ferramentas similares ao RECALL. 

%% file: figuras/contratoRCLErrado.tex
\centering
\noindent
\begin{minipage}{\textwidth} 
	\begin{framed} 
		\[
		\begin{array}{r @{\hspace{1em}} l}
			1 & \{b,s\} [\text{buyProduct}] ( \\
			2 & \quad \{b,k\} O (\text{payProduct})^\land \\
			3 & \quad \{b,k\} [\text{payProduct}] ( \\
			4 & \quad \quad \{k,s\} O (\text{notifyProductPayment})^\land \\
			5 & \quad \quad \{k,s\} [\text{notifyProductPayment}] ( \\
			6 & \quad \quad \quad \{s,c\} O (\text{sendProduct})^\land \\
			7 & \quad \quad \quad \{s,k\} O (\text{payShippingCosts})^\land \\
			8 & \quad \quad \{s,k\} [\text{payShippingCosts}] ( \\
			9 & \quad \quad \quad \quad \{s,c\} [\text{sendProduct}] ( \\
			10 & \quad \quad \quad \quad \quad \{c,b\} O (\text{deliverProduct})^\land \\
			11 & \quad \quad \quad \quad \quad \{c,b\} [\text{deliverProduct}] ( \\
			12 & \quad \quad \quad \quad \quad \quad \{b,k\} O (\text{notifyProductReceipt})^\land \\
			13 & \quad \quad \quad \quad \quad \quad \{c,s\} O (\text{notifyProductDelivery}) \\
			14 & \quad \quad \quad \quad \quad \quad \{b,k\} [\text{notifyProductReceipt}] (\{k,s\} O (\text{payProduct}))^\land \\
			15 & \quad \quad \quad \quad \quad \quad \{c,s\} [\text{notifyProductDelivery}] ( \\
			16 & \quad \quad \quad \quad \quad \quad \quad \{s,k\} O (\text{liberateShippingCosts})^\land \\
			17 & \quad \quad \quad \quad \quad \quad \quad \{s,k\} [\text{liberateShippingCosts}] \\
			18 & \quad \quad \quad \quad \quad \quad \quad \quad (\{k,c\} O (\text{payShippingCosts})) \\
			19 & \quad \quad \quad \quad \quad \quad \quad \quad )))))))))); \\
			20 & \{b,k\} [\neg \text{notifyDelivery}] * (\{k,s\} \mathcal{F} (\text{payProduct})); \\
			21 & \{s,k\} [\neg \text{liberateShippingCosts}] * (\{k,c\} \mathcal{F} (\text{payShippingCosts})); \\
			22 & \{s,c\} [\neg \text{payShippingCosts}] * (\{c,b\} \mathcal{F} (\text{deliverProduct})); \\
		\end{array}
		\]
	\end{framed}
	\caption{Contrato de compra e venda em RCL}
	\label{fig:rclerradoteste}
\end{minipage}

%% file: figuras/contratoRCLCorreto.tex
\centering
\noindent
\begin{minipage}{\textwidth}
	\begin{framed}
		\[
		\begin{array}{r @{\hspace{1em}} l} 
			9 & \quad \{k,c\} O(\text{notifyShippingPayment}) \land \{s,c\} [\text{sendProduct}] ( \\
			21 & \{k,c\} [(\text{notifyShippingPayment})*] (\{c,b\} \mathcal{F}(\text{deliverProduct})); \\
		\end{array}
		\]
	\end{framed}
	\caption{Trecho do contrato corrigido em RCL}
	\label{fig:rclcorretoteste}
\end{minipage}

%% file: rcl-solidity.tex
\section{Uma Abordagem de Verificação para Contratos Inteligentes}\label{aplicacao-solidity}

Uma abordagem rigorosa de verificação para contratos inteligentes deve, invariavelmente, recair sobre métodos e abordagens que se utilizam de formalismos e embasamento matemático. 
Uma possível abordagem de verificação mais rigorosa pode se valer da precisão da RCL. 
No entanto, existe uma lacuna entre a linguagem Solidity, de alto nível que descreve contratos inteligentes, e a RCL, geralmente usada para descrever contratos eletrônicos. 
Uma proposta para transpor essa lacuna é uma transformação em múltiplas etapas que garanta a mesma expressividade e propriedades do contrato RCL em Solidity. 

O teste de conceito proposto neste trabalho consiste na implementação em Solidity do contrato apresentado na Subseção~\ref{sec:contratorcl-conflito}.  
A obtenção do contrato inteligente que implementa a compra e venda de produtos tal como descrito anteriormente é concebido com base nas 
cláusulas em RCL. 
Essas regras descritas nas cláusulas do contrato são implementadas como funções em Solidity. 
O processo de obtenção do contrato em Solidity a partir da RCL é o mesmo adotado tanto para o contrato contendo conflito quanto para o contrato corrigido. 

A Subseção~\ref{sec:contratosol-conflito} mostra como o contrato descrito na subseção~\ref{sec:contratorcl-conflito} é implementado em Solidity, especificando as obrigações, permissões e proibições através de um programa executável para uma \emph{blockchain}. 
O contrato é testado a fim de garantir que a semântica em RCL seja traduzida de forma apropriada para contratos inteligentes em Solidity. 
O mesmo processo de transformação é então replicado para a versão corrigida do RCL,  na Subseção~\ref{sec:contratosol-corrigida}, mostrando uma execução correta do contrato. 

Este estudo possibilita uma prova de conceito sobre o uso de ferramentas de verificação através da transformação de contratos implementados em Solidity e suas respectivas descrições em RCL, além de  permitir uma análise sobre os desafios e as potenciais perdas de informações que impactam essa transformação, reforçando a necessidade de uma abordagem sistemática para garantir a corretude e a segurança de contratos inteligentes. 


\subsection{Implementação do Contrato em Solidity}\label{sec:contratosol-conflito}

O código Solidity para a primeira versão do contrato de compra e venda, que contém um conflito normativo, é obtido pela tradução manual a partir do contrato em RCL para Solidity. 
Logo, o contrato em Solidity também possui quatro agentes: 
o comprador, o vendedor, o banco e a transportadora; 
identificados pelos endereços de suas respectivas contas no Ethereum. 
Além disso, o contrato possui estados que modelam seu fluxo de controle, simulando a execução na ordem correta das cláusulas em RCL. 
Os estados de controle criados no contrato inteligente são \emph{Created}, \emph{ProductBought}, \emph{ProductPaid}, \emph{PaymentNotified}, \emph{ProductDelivered} e \emph{Finalized}, 
e representam, respectivamente, a criação do contrato, a compra 
de um produto, o pagamento do produto realizado pelo comprador, a notificação de pagamento da compra realizada, 
a entrega do produto, e a finalização do contrato. 

A passagem do controle de um estado para outro indica as mudanças e ações que representam as cláusulas do contrato, bem como as regras internas do banco e da transportadora. 
Algumas variáveis também servem para controlar as ações concorrentes do contrato, como por exemplo, as ações realizadas 
após o pagamento do produto ser notificado: 
(a) enviar o produto; e (b) pagar o frete. 
Após essas duas ações, especificadas por duas variáveis 
booleanas, serem concluídas, a próxima ação pode então ser executada. 

Outras variáveis de controle cumprem o mesmo papel no contrato: \emph{receiptNotifiedByBuyer, deliveryNotifiedByCarrier, paymentReleasedSeller, paymentReleasedCarrier}. 
Já a regra interna da transportadora é traduzida para a variável booleana \emph{paymentReleasedCarrierSeller}, enquanto que a regra interna do banco é especificada pela variável \emph{receiptNotifiedByBuyer}, que muda seu valor verdade quando o comprador notifica o recebimento do produto 
para que o banco possa liberar o valor da venda para o vendedor, 
e pela função \emph{liberateShippingCosts}, que realiza 
o pagamento do frete para a transportadora caso nenhum erro ocorra. 

O código também define os eventos que notificam as mudanças de estado e os modificadores que verificam o agente 
responsável pela ação: onlyB, onlyS, onlyC e onlyK. 
O modificador onlyB, por exemplo, verifica 
o responsável pela ação do agente que está realizando a compra através do seu endereço. 
Caso não seja do comprador, a função não é realizada, 
garantindo a característica da relativização modelada pela 
RCL, onde cada ação tem um participante responsável associado. 
O modificador pode ser observado na função \emph{buyProduct}, por exemplo, onde apenas o usuário que 
possui o endereço específico de comprador tem a permissão de realizar a compra do produto. 

O construtor do contrato, uma função especial executada apenas uma vez na implantação do contrato na \emph{blockchain}, é a responsável por receber informações ou instruções que irão inicializar o contrato, bem como o construtor de uma classe. No caso desse contrato, ele recebe as informações das contas das partes, e dos valores de frete e produto, por meio de seis parâmetros:
\begin{itemize}
	\item os endereços Ethereum (\emph{address}) dos quatro participantes: \_buyer, \_seller, \_bank e \_carrier; 
	\item o valor total do pagamento (\_paymentAmount);
	\item o valor do frete.
\end{itemize}
Estas informações são armazenadas nas variáveis de estado do contrato, declaradas no início do código, e permanecem imutáveis durante toda a vida do contrato. 
O construtor também inicializa o estado do contrato como \emph{Created}, marcando o início do processo de compra e venda.

As funções implementadas especificam as cláusulas do contrato, dispostas na mesma ordem em que aparecem no contrato descrito anteriormente. 
Os modificadores garantem que as funções sejam executadas somente mediante certas condições. 
Os modificadores \emph{only} garantem que certas ações só podem ser executadas por seus respectivos participantes. 
O modificador \emph{onlyB}, por exemplo, permite apenas ações executadas pelo \emph{buyer}. 
Já os modificadores \emph{internalRules} são usados para descrever as regras internas do contrato.

O contrato de compra e venda, escrito em Solidity e apresentado na Listagem~\ref{sol-compra-venda-conflito}, 
é obtido da versão RCL original. 

\input{figuras/rcl-original}

O contrato é implantado e executado no \emph{Ethereum} pela plataforma Remix IDE, como segue: 
\begin{enumerate}
	\item o contrato é criado com os respectivos endereços de cada participante, com um valor (um inteiro qualquer) atribuído ao produto da venda e 
	um valor (um inteiro qualquer) associado ao frete; 
	
	\item a função \emph{buyProduct} é executada pela conta do buyer, mudando o estado do contrato de \emph{Created} para \emph{ProductBought}; 
	
	\item após o compra do produto a função \emph{payProduct} também é executada pelo buyer e muda o estado do contrato de \emph{Created} para \emph{ProductPaid}; 
	
	\item o banco então notifica o pagamento do produto através da função \emph{notifyProductPayment}. 
	Se a função é executada com sucesso, o controle do contrato passa para 
	o estado \emph{PaymentNotified}; 
	
	\item na sequência o seller executa duas funções, \emph{sendProduct} e \emph{payShippingCosts}. 
	Se as duas funções são executadas com sucesso, é possível seguir no contrato e executar a próxima função, no caso \emph{deliverProduct};
	
	\item em seguida, a função \emph{deliverProduct} é executada. 
	Porém, a mensagem de erro \emph{``Frete não foi pago pelo vendedor a transportadora''} é emitida, como esperado, já que essa é a regra interna da transportadora. 
	Essa regra impede a execução correta do contrato devido ao conflito entre a cláusula que exige o envio do produto antes da entrega e cláusula que exige o pagamento do produto antes que esse seja enviado. 
	Porém, apenas uma delas é satisfeita até este ponto da execução do contrato, ou seja, o frete ter sido pago. 
	Logo, a função não pode ser executada corretamente, passando o controle da execução do contrato para o próximo estado, travando a execução do contrato neste estado.  

\end{enumerate}

\subsection{Versão Corrigida em Solidity}\label{sec:contratosol-corrigida}

Assim como a versão original em RCL é transformada em código Solidity, a versão corrigida em RCL, apresentada na Seção~\ref{sec:contratorcl-corrigido}, também é traduzida para código Solidity. 
A modificação consiste em o banco notificar a transportadora quando o pagamento do frete é efetuado pelo vendedor. 
Dessa forma, a transportadora considera essa notificação do banco como garantia para liberar a entrega do produto ao comprador. 

O contrato em Solidity resultante da transformação possui então algumas diferenças com relação a primeira versão. 
A variável \emph{paymentRealeasedCarrierSeller}, que modelava a antiga regra da transportadora, é alterada para \emph{shippingPaymentNotified}. 
A mudança do nome da variável, claramente, não resolve o conflito em si, mas mantém um alinhamento com a nova regra da transportadora. 
O conflito, propriamente dito, é resolvido com a adição de 
uma nova função, chamada \emph{notifyShippingPaymentToCarrier}, que 
é executada antes do produto ser enviado ao comprador. 
Essa função reflete a descrição da linha~9 do contrato RCL corrigido, onde 
após o vendedor pagar a taxa do frete para o banco, este deve informar a transportadora da realização do pagamento (cláusula \emph{notifyShippingPayment} em RCL), para que a transportadora possa enviar o produto. 
Uma última modificação é realizada na função \emph{deliverProduct}, avaliando as condições de satisfação através das variáveis \emph{productSent} e \emph{shippingPaymentNotified}. 
Quando as condições são verdadeiras o fluxo de execução do contrato ocorre corretamente, evitando o erro com a variável \emph{paymentRealeasedCarrierSeller}. 
O trecho em Solidity com as 
alterações propostas estão descritas na Listagem~\ref{sol-compra-venda-corrigido}. 
\input{figuras/rcl-modif}

Os testes realizados com a nova versão corrigida do contrato em Solidity diferem dos originais nos seguintes passos: 
\begin{enumerate}
	\item[5.] Após o pagamento do produto, o seller executa duas funções,  \emph{sendProduct} e \emph{payShippingCosts}, e o banco executa a função \emph{notifyShippingPaymentToCarrier}. 
	Se todas as funções são executadas com sucesso, o contrato segue normalmente e executa a próxima função, \emph{deliverProduct};
\end{enumerate}

Com a nova versão do contrato em Solidity, novos testes no Remix IDE indicaram uma execução correta, diferente da versão apresentada na Subseção~\ref{sec:contratosol-conflito}, conforme passo~6 da execução original. Na execução do primeiro contrato, era possível executar corretamente as funções \emph{buyProduct, payProduct e notifyProductPayment, sendProduct e payShippingCosts}, porém o contrato ficava travado na função \emph{deliverProduct}. Nesse contrato o resultado é que o estado do contrato consegue chegar até \emph{shippingCostsPaid}, porém não é possível avançar para o final do contrato (no caso, para o estado \emph{ProductDelivered}). Já na execução com o contrato corrigido, é possível executar todas as funções sem erros e impedimentos, tendo ao final, um contrato com o estado \emph{ProductDelivered}, que indica o estado final do contrato.

%% file: figuras/rcl-original.tex
\begin{lstlisting}[language=Solidity, basicstyle=\scriptsize\ttfamily,caption=Código Solidity do contrato com conflito,label={sol-compra-venda-conflito}]
// SPDX-License-Identifier: MIT
pragma solidity ^0.8.0;

contract ContratoComErro {
	address public buyer;
	address public seller;
	address public bank;
	address public carrier;
	uint public paymentAmount;
	uint public shippingCosts;
	
	enum ContractState {
		Created,
		ProductBought,
		ProductPaid,
		PaymentNotified,
		ProductDelivered,
		Finalized
	}
	ContractState public state;
	
	bool private productSent = false; // {s,c}O(sendProduct)
	bool private shippingCostsPaid = false; // {s,k}O(payShippingCosts)
	bool private receiptNotifiedByBuyer = false; // {b,k}O(notifyProductReceipt)
	bool private deliveryNotifiedByCarrier = false; // {c,s}O(notifyProductDelivery)
	bool private paymentReleasedSeller = false;
	bool private paymentReleasedCarrier = false;
	bool private paymentRealeasedCarrierSeller = false;
	
	event Notify(
	address indexed sender,
	address indexed receiver,
	string message
	);
	
	modifier onlyB() {
		require(msg.sender == buyer, "Apenas o Comprador (b)");
		_;
	}
	modifier onlyS() {
		require(msg.sender == seller, "Apenas o Vendedor (s)");
		_;
	}
	modifier onlyK() {
		require(msg.sender == bank, "Apenas o Banco (k)");
		_;
	}
	modifier onlyC() {
		require(msg.sender == carrier, "Apenas a Transportadora (c)");
		_;
	}
	
	constructor(
	address _buyer,
	address _seller,
	address _bank,
	address _carrier,
	uint _paymentAmount,
	uint _shippingCosts
	) {
		buyer = _buyer;
		seller = _seller;
		bank = _bank;
		carrier = _carrier;
		paymentAmount = _paymentAmount;
		shippingCosts = _shippingCosts;
		state = ContractState.Created;
	}
	
	modifier atState(ContractState _requiredState) {
		require(state == _requiredState, "Estado invalido para essa acao");
		_;
	}
	
	// 1. {b,s}[buyProduct](...)
	function buyProduct() external onlyB atState(ContractState.Created) {
		state = ContractState.ProductBought;
		emit Notify(buyer, seller, "1. Comprador realizou a compra.");
	}
	
	// 2. {b,k}O(payProduct)
	function payProduct()
	external
	payable
	onlyB
	atState(ContractState.ProductBought)
	{
		require(msg.value == paymentAmount, "Valor do pagamento incorreto");
		state = ContractState.ProductPaid;
		emit Notify(buyer, bank, "2. Comprador pagou o produto ao banco.");
	}
	
	// 4. {k,s}O(notifyProductPayment)
	function notifyProductPayment()
	external
	onlyK
	atState(ContractState.ProductPaid)
	{
		state = ContractState.PaymentNotified;
		emit Notify(
		bank,
		seller,
		"4. Banco notificou o vendedor sobre o pagamento."
		);
	}
	
	// 6. {s,c}O(sendProduct)
	function sendProduct()
	external
	onlyS
	atState(ContractState.PaymentNotified)
	{
		require(!productSent, "Produto ja foi enviado");
		productSent = true;
		emit Notify(
		seller,
		carrier,
		"6. Vendedor enviou o produto para a transportadora."
		);
	}
	
	// 7. {s,k}O(payShippingCosts)
	function payShippingCosts()
	external
	payable
	onlyS
	atState(ContractState.PaymentNotified)
	{
		require(msg.value == shippingCosts, "Valor do frete incorreto");
		require(!shippingCostsPaid, "Frete ja foi pago");
		shippingCostsPaid = true;
		emit Notify(seller, bank, "7. Vendedor pagou o frete ao banco.");
	}
	
	// 10. {c,b}O(deliverProduct)
	// 21. {s,c}[(!payShippingCosts)*]({c,b}F(deliverProduct))
	function deliverProduct() external onlyC {
		require(productSent, "Produto ainda nao foi enviado pelo vendedor");
		
		require(
		shippingCostsPaid,
		"ERRO : Transportadora nao pode entregar antes do frete ser pago."
		);
		
		require(
		paymentRealeasedCarrierSeller,
		"Frete nao foi pago pelo vendedor a transportadora"
		);
		
		state = ContractState.ProductDelivered;
		emit Notify(
		carrier,
		buyer,
		"10. Transportadora entregou o produto ao comprador."
		);
	}
	
	// 12. {b,k}O(notifyProductReceipt)
	function notifyProductReceipt()
	external
	onlyB
	atState(ContractState.ProductDelivered)
	{
		require(!receiptNotifiedByBuyer, "Recebimento ja foi notificado");
		receiptNotifiedByBuyer = true;
		emit Notify(
		buyer,
		bank,
		"12. Comprador notificou o banco do recebimento."
		);
	}
	
	// 13. {c,s}O(notifyProductDelivery)
	function notifyProductDelivery()
	external
	onlyC
	atState(ContractState.ProductDelivered)
	{
		require(!deliveryNotifiedByCarrier, "Entrega ja foi notificada");
		deliveryNotifiedByCarrier = true;
		emit Notify(
		carrier,
		seller,
		"13. Transportadora notificou o vendedor da entrega."
		);
	}
	
	// 14. {b,k}[notifyProductReceipt]({k,s}O(payProduct))
	// 19. Regra Interna do Banco
	function payProductSeller()
	external
	onlyK
	atState(ContractState.ProductDelivered)
	{
		require(
		receiptNotifiedByBuyer,
		"Regra Interna B: Comprador ainda nao confirmou o recebimento."
		);
		
		emit Notify(bank, seller, "14. Banco liberou o pagamento ao vendedor.");
		paymentReleasedSeller = true;
		checkFinalization();
	}
	
	// 16. {s,k}O(liberateShippingCosts)
	function liberateShippingCosts()
	external
	onlyS
	atState(ContractState.ProductDelivered)
	{
		require(
		deliveryNotifiedByCarrier,
		"Vendedor ainda nao foi notificado pela transportadora."
		);
		
		emit Notify(
		seller,
		bank,
		"16. Vendedor autorizou banco a liberar frete."
		);
		payShippingCostsToCarrier();
	}
	
	// 17. {k,c}O(payShippingCosts)
	// 20. Regra Interna do Banco
	function payShippingCostsToCarrier() private onlyK {
		emit Notify(bank, carrier, "17. Banco pagou o frete a transportadora.");
		paymentReleasedCarrier = true;
		checkFinalization();
	}
	
	function checkFinalization() private {
		if (state == ContractState.ProductDelivered) {
			if (paymentReleasedCarrier && paymentReleasedSeller) {
				state = ContractState.Finalized;
			}
		}
	}
}	
\end{lstlisting}

%% file: figuras/rcl-modif.tex
\begin{lstlisting}[language=Solidity, basicstyle=\scriptsize\ttfamily, caption=Techo do código Solidity livre de conflito, label={sol-compra-venda-corrigido}]
	
	function notifyShippingPaymentToCarrier()
	external
	onlyK
	atState(ContractState.PaymentNotified)
	{
		require(
		shippingCostsPaid,
		"O vendedor ainda nao pagou o frete ao banco."
		);
		require(
		!shippingPaymentNotified,
		"Notificacao de frete ja foi enviada."
		);
		
		shippingPaymentNotified = true;
		emit Notify(
		bank,
		carrier,
		"9. Banco notificou a transportadora sobre o pagamento do frete."
		);
	}
	
	function deliverProduct() external onlyC {
		require(productSent, "Produto ainda nao foi enviado pelo vendedor");
		
		require(
		shippingPaymentNotified,
		"Regra Interna C: Transportadora so pode entregar apos NOTIFICACAO do banco."
		);
		
		state = ContractState.ProductDelivered;
		emit Notify(
		carrier,
		buyer,
		"10. Transportadora entregou o produto ao comprador."
		);
	}
\end{lstlisting}

%% file: conclusao.tex
\section{Conclusão}

Este trabalho reforça a importância na adoção de métodos formais no ciclo de desenvolvimento de aplicações práticas. 
A utilização da Linguagem de Contrato Relativizada (RCL) como instrumento de especificação para contratos normativos mostrou-se uma alternativa promissora para superar limitações presentes em formalismos tradicionais, especialmente no que diz respeito à representação explícita das responsabilidades e relações entre as partes envolvidas de contratos multilaterais.

O emprego da ferramenta RECALL, na verificação de cláusulas contratuais descritas em RCL para a detecção de conflitos normativos, permitiu uma prova de conceito para a  mitigação de erros na implementação de contratos inteligentes em Solidity. 
A abordagem mostra que é possível reduzir significativamente o risco de erros lógicos que, em ambientes baseados em \emph{blockchain}, podem resultar em prejuízos financeiros e jurídicos irreversíveis. 

O trabalho permitiu a modelagem de um cenário multilateral, identificando explicitamente as obrigações e proibições de cada participante do contrato. 
A abordagem proposta então demonstra que a verificação formal não deve ser vista apenas como uma etapa posterior de validação, mas como um componente essencial do processo de engenharia de contratos inteligentes. 
Ao antecipar a análise de propriedades normativas, tais como obrigações, permissões e proibições associadas a agentes específicos, cria-se um ambiente mais propício à construção de sistemas descentralizados seguros, transparentes e auditáveis.



Os experimentos realizados no ambiente Remix IDE reforçam a viabilidade prática dessa integração entre especificação formal e implementação, mostrando que a tradução ad-hoc de contratos verificados em RCL para Solidity, embora ainda dependa de intervenção manual, pode ser guiada de maneira sistemática pelos resultados da verificação. 
Esse aspecto evidencia uma oportunidade concreta para a automação futura desse processo, por meio do desenvolvimento de algoritmos ou tradutores que conectem diretamente linguagens formais normativas a linguagens de programação voltadas a \emph{Blockchains}.

As principais contribuições deste trabalho são: 
(i) a aplicação da RCL como mecanismo de verificação prévia de contratos inteligentes; 
(ii) a demonstração do papel da ferramenta RECALL como suporte à análise de conflitos normativos em cenários multilaterais; e 
(iii) a proposição de um fluxo sistemático que possa integrar a verificação formal e a implementação prática de contratos inteligentes. 
Tais contribuições reforçam a relevância da pesquisa no contexto atual, em que a confiabilidade de sistemas descentralizados é cada vez mais crítica.


Como direções futuras para esse trabalho, vislumbra-se a ampliação desta abordagem para diferentes contratos inteligentes, bem como a investigação de mecanismos para a tradução e geração automática de código Solidity a partir de especificações em RCL. 
Espera-se que esses avanços possam consolidar ainda mais a adoção de técnicas formais no ecossistema de \emph{blockchain}, promovendo o desenvolvimento de contratos inteligentes mais seguros, corretos e alinhados às exigências normativas e legais dos ambientes digitais contemporâneos.